\documentclass[12pt]{article}

\usepackage{amssymb}
\usepackage{amsmath}
\usepackage{latexsym}
\usepackage{yfonts}

\catcode`@=11
\renewcommand{\section}{\@startsection{section}{1}{0pt}{\medskipamount}
{\medskipamount}{\large\bf}} \numberwithin{equation}{section}
\catcode`@=12

\def\beq{\begin{eqnarray}}    
\def\eeq{\end{eqnarray}}      

\def\={\ =\ }

\begin{document}

\begin{center}

{\Large\bf Superfield generating equation of field-antifield
formalism as a hyper-gauge theory
}

\vspace{18mm}

{\large Igor A. Batalin$^{(a,b)}\footnote{E-mail:
batalin@lpi.ru}$\;, Peter M. Lavrov$^{(b, c)}\footnote{E-mail:
lavrov@tspu.edu.ru}$\; }

\vspace{8mm}

\noindent ${{}^{(a)}}$
{\em P.N. Lebedev Physics Institute,\\
Leninsky Prospect \ 53, 119991 Moscow, Russia}

\noindent  ${{}^{(b)}} ${\em
Tomsk State Pedagogical University,\\
Kievskaya St.\ 60, 634061 Tomsk, Russia}

\noindent  ${{}^{(c)}} ${\em
National Research Tomsk State  University,\\
Lenin Av.\ 36, 634050 Tomsk, Russia}

\vspace{20mm}

\begin{abstract}
\noindent
Within  a superfield  approach,  we formulate a  simple quantum generating  equation
of  the field-antifield  formalism.  Then we  derive  the  Schroedinger  equation
with the  Hamiltonian  whose $\Delta$-exact  part serves as a generator
to  the quantum master-transformations. We show that these generators
do satisfy a nice composition law in terms of  the quantum antibrackets.
We also present an $Sp(2)$ symmetric extension to the main construction,
with specific features caused by the principal fact that all basic equations become
$Sp(2)$ vector-valued ones.
\end{abstract}

\end{center}

\vfill

\\

\newpage

\section{Introduction}

From the early days of the field-antifield formalism, a fundamental idea
was presented \cite{BV,BV1} as to how to formulate a universal
hyper-gauge theory whose gauge generators
would, by construction,  be included naturally into the Hessian of the original master
action of the universal theory, defined so as to satisfy the (classical) master equation
formulated in terms of  the antibrackests \cite{Schout,Butt}.
Then the notion of a proper solution to
the master equation was defined by requiring that there were no other gauge generators
involved than the ones included into the Hessian.
The next step was made by formulating
the quantum master equations in terms of the odd Laplacian operator. The quantum
master equation was derived later directly from the Hamiltonian formalism
\cite{BF-Poin,GGT}.
These basic ideas were developed as a success \cite{BV,BV1},
as applied to both the irreducible,
and  to the reducible gauge theories.  In general,
the universal hyper-gauge
theory was invented so as to "be ready" to include into itself any
possible particular model
with a gauge-invariant initial action.

In  the present paper, we develop further the profound idea of the
hyper-gauge theory at the quantum superfield level. Firstly, within
a superfield approach \cite{BBD1,BBD2,BL-IJMP}, we formulate a
simple quantum generating equation of the field-antifield formalism
as having its configuration space identified with the antisymplectic
phase space of fields and antifields. The latter generating equation
is presented in terms of a superfield covariant derivative with
respect to the two-dimensional super-time whose Boson component is
the "ordinary" time, purely formal in its origin, while its Fermion
component is identified
naturally with the BRST parameter. The covariant derivative squared
is just the "ordinary" time derivative. Then we derive the standard
Schroedinger equation by applying again the covariant derivative to
the  generating superfield equation. We provide effectively for the
Hamiltonian commuting with the odd Laplacian ( the $\Delta$
operator).  As usual, the Hamiltonian consists of a singlet part and
a $\Delta$-exact  part. In particular,  in the absence of a
singlet component,  the Hamiltonian becomes purely $\Delta$-exact.
We show that the $\Delta$-exact part of the Hamiltonian serves as
a generator to the quantum master-transformations.  Classically,
these transformations consist of the two pieces: the first of them
is just an anticanonical transformation, while the second is caused
by the Jacobian of the transformation. Then we show  that the
generators of the quantum master transformations do satisfy a very
nice composition law as formulated in terms of the so-called quantum
antibrackets \cite{BM1,BM2}. We  also present an $Sp(2)$ symmetric
extension  to the main construction, with specific features caused
by the principal fact that all basic equations become $Sp(2)$
vector-valued ones.

\section{Superfield generating equation}

It appears to be a remarkable feature that the generating equation of the
field-antifield formalism takes the very simple form of a superfield
Schroedinger equation,
\beq
\label{SFBV2}
(  i \hbar D  -  Q )
\Psi =  0,\quad
D  =:   \frac{ \partial }{ \partial \tau }  +
\tau   \frac{ \partial }{ \partial t }, \quad Q=: \Delta - F,   
\eeq where $D$ is a covariant super-time derivative, $Q$ is a
super-charge whose kinetic part is the odd Laplacian, $\Delta$, and
$F$ is a super-potential depending on momenta in general,
\beq
\label{SFBV3} &&\varepsilon( D )  =
1,  \quad D^{ 2 }  =
\frac{1}{ 2 }  [ D, D ]  = \frac{ \partial }{ \partial t }\;\!, \\ 
\label{SFBV3a}
&&\;\varepsilon( \Delta )  =  1, \quad \Delta^{ 2 }  =
\frac{ 1 }{ 2 }  [ \Delta, \Delta ]  =  0, \\   
\label{SFBV4}
&&\quad F   =  :  F( Z, P ), \quad \varepsilon( F )  =  1.    
\eeq
The equation (\ref{SFBV2}) is formulated for a  superfield,
\beq
\label{SFBV1}
\Psi  =:  \Psi( t, \tau, Z ),
\quad  \varepsilon( t )  =  0, \quad \varepsilon( \tau )  =  1. 
\eeq
We assume that the co-ordinate operators $Z^{ A }$ are identified with
the standard full set of the field-antifield
variables, and $P_{ A }$ are their respective canonically-conjugate momenta operators,
\beq
\label{SFBV6a}
[ Z^{ A },  Z^{ B } ]  =  0,  \quad  [ Z^{ A },  P_{ B } ]  =  i \hbar  \delta^{ A }_{ B },
\quad    [ P_{A},  P_{ B } ]  =  0 .    
\eeq
It follows from (\ref{SFBV2}) that the standard Schroedinger equation holds
\beq
\label{SFBV5}
i  \hbar  \frac{ \partial }{ \partial t } \Psi  =  \mathcal{ H }  \Psi,     
\eeq
with the Hamiltonian
\beq
\label{SFBV6}
 \mathcal{ H }  =:  -(i \hbar)^{-1} \frac{1}{2}[Q,Q]=( i \hbar )^{ -1}
 [  (  \Delta  -  \frac{ 1 }{ 2 } F  ) ,  F  ].  
\eeq
The superfield (\ref{SFBV1}) has the component form
\beq
\label{SFBV6a}
\Psi( t, \tau, Z )  =  \left(  1  +  \tau  ( i \hbar ) ^{-1}  Q  \right)
\Psi_{ 0 }( t, Z ),    
\eeq
where the zero-component $\Psi_{ 0 }( t, Z )$ satisfies  by itself
the equation (\ref{SFBV5}) with the Hamiltonian (\ref{SFBV6}).
As for an arbitrary $F$, the Hamiltonian (\ref{SFBV6})
does not commute with the $\Delta$.
However, it follows from (\ref{SFBV6}) that
\beq
\label{SFBV7a}
[ Q,  \mathcal{ H } ]  =  0.  
\eeq
Thus, we arrive at the implication
\beq
\label{SFBV7b}
[ \Delta, \mathcal{ H } ]  =  0 \; \Rightarrow \;
[ \mathcal{ H },  F ]  =  0,    
\eeq
or more explicitly
\beq
\label{SFBVc}
[ [ \Delta, F ], F ]  =  [ \Delta, \frac{1}{2} [ F, F ] ]  =  0.   
\eeq
Due to the Poincare lemma, we have
\beq
\label{SFBV7d}
\frac{1}{2} [ F, F ]  = - i \hbar \mathcal{ H }_{ S }  -  [ \Delta,  G ],   
\eeq
where $\mathcal{ H }_{ S }$ is a Boson singlet component,
\beq
\label{SFBV7e}
[ \Delta,  \mathcal{ H }_{ S } ]  =  0, \quad
\mathcal{ H }_{ S }   \neq [ \Delta,  {\rm anything} ],   
\eeq
$G$ is an arbitrary Fermion operator.
By inserting (\ref{SFBV7d}) into (\ref{SFBV6}), we get
\beq
\label{SFBV7f}
\mathcal{ H }  =  \mathcal{ H }_{ S }  +  \mathcal{ H }_{ \Phi },  
\eeq
where the $\Delta$-exact $\Phi$ component is defined as
\beq
\label{SFBV7g}
\mathcal{ H }_{ \Phi }   =:  ( i \hbar )^{-1} [ \Delta,  \Phi  ],\quad
\Phi   =:  F  +  G.   
\eeq
As the $G$ in the second in (\ref{SFBV7g}) is an
arbitrary Fermion operator, the respective natural arbitrariness
is inherited in (\ref{SFBV7f}), as well, with having the implicit
$G$ dependence  taken into  account in the  $F$,  via the equation
(\ref{SFBV7d})
with the singlet component $\mathcal{ H }_{S}$ being  kept fixed.
In its turn, the equation (\ref{SFBV7d})
rewrites in the equivalent form,
\beq
\label{SFBV7ga}
\frac{ 1 }{ 2 }  (  i  \hbar )^{-1}  (  [  G,  G  ]  -  [  \Phi,  \Phi  ]  )  =
\mathcal{ H }_{ S }  +  ( i \hbar )^{ -1 }  [  Q,  G  ].    
\eeq

Once the $\Delta$ operator commutes with the Hamiltonian $\mathcal{ H }$,
it follows from the (\ref{SFBV5}) for the zero component $\Psi_{0}$
\beq
\label{SFBV7h}
i \hbar \frac{\partial}{\partial t } \Delta \Psi_{0}  =  \mathcal{ H }  \Delta \Psi_{0},    
\eeq
\beq
\label{SFBV7j}
\Delta \Psi_{0} |_{ t =  0 }  =  0  \;\Rightarrow \;
\Delta \Psi_{0} |_{ {\rm any}\;\! t }  =  0.  
\eeq
The implication (\ref{SFBV7j}) shows that the
arbitrariness of a solution to the quantum
master equation,
\beq
\label{SFBV7k}
\Delta \Psi_{ 0}  =  0,  \quad \varepsilon(\Psi_{ 0})=0,\quad
\Psi_{ 0 }  =:  \exp\left\{ \frac{ i }{ \hbar } W \right\},   
\eeq
is measured by the evolution operator,
\beq
\label{SFBV7m}
\Psi_{ 0 }|_{t=0} \; \rightarrow \;
\Psi_{0}|_{any\;t}=
\exp\left\{ - \frac{ i }{ \hbar }\mathcal{ H }\;\!t \right\}\Psi_{ 0 }|_{t=0}. 
\eeq

\section{Quantum master-transformations and their composition law}

Now, consider a family of operators
\beq
\label{SFBV28}
\mathcal{ H }_{ F }  =:  ( i \hbar)^{ -1}  [ \Delta,  F ],      
\eeq
with $F( Z, P )$ being an arbitrary Fermion operator.
By definition, the equation (\ref{SFBV28})
is a generator of a quantum master-transformation \cite{BLT-EPJC}.
Notice that the operator (\ref{SFBV28}) can be rewritten naturally in terms of
both the free-acting operators $P_A$ and the adjoint-acting ones $P'_A$,
\beq
\label{SFBV29}
i \hbar  \mathcal{ H }_{F}  =  ( \Delta' F )  -  {\rm ad}' ( F ), 
\eeq
where we have used the definitions
\beq
\label{SFBV30}
{\rm ad}'( F )  =:
( F \overleftarrow{P'}_{ A } )E^{ AB } P_{ B } (-1)^{\varepsilon_{ B } },   
\eeq
\beq
\label{SFBV18}
\Delta  =:  \frac{ 1 }{ 2 } P_{ A } E^{ AB } P_{ B } (-1)^{ \varepsilon_{ B } }, \quad
E^{ AB }  =  {\rm const}, \quad
\Delta'  =:  \Delta\big|_{ P \rightarrow P'}\;,     
\eeq
\beq
\label{SFBV19}
P_{ A }  =:  - i \hbar  \partial_{ A }  (-1)^{ \varepsilon_{ A } }, \quad
 P'_{ A }  =:  {\rm ad}( P_{ A } ) =  [ P_{ A }, \; \cdot \; ]  ,\quad   
 \overleftarrow{ P' }_{\! A }  =:  -  [\;\cdot\; ,  P_{ A }  ].
\eeq
Due to the Jacobi identity for (super)commutators,
the following relations hold for arbitrary operators $A, B$,
\beq
\label{SFBV19a}
{\rm ad}(A)=: [A,\;\cdot\;]\quad \Rightarrow \quad
[{\rm ad}(A), {\rm ad}(B)]={\rm ad}([A,B]).
\eeq

 From the classical point of view , in the right-hand side
in (\ref{SFBV29}), the second term describes an anticanonical
transformation with $F$ being a generator, while the first term is
caused by the Jacobian of the latter transformation.

A solution to the Schroedinger equation (\ref{SFBV5}) with the Hamiltonian
(\ref{SFBV29}) has the form of a quantum anticanonical transformation,
\beq
\label{SFBV30a}
\Psi  =  \exp\left\{  -  ( i \hbar )^{ -2 } \;\! t\;  {\rm ad}'( F )\right\}\Psi_{ J },
\eeq
where the  "Jacobian wave function",  $\Psi_{ J }$,  does satisfy the
equation
\beq
\label{SFBV30b}
\partial_{ t }  \Psi_{ J }  =
\exp\left\{  ( i \hbar )^{ -2 } \;\! t \; {\rm ad}'( F )  \right\}
( i \hbar )^{ -2 }  (  \Delta' F  )
\exp\left\{  -  ( i \hbar )^{ -2 } \;\! t\;  {\rm ad}'( F )  \right\}  \Psi_{ J }.   
\eeq
In the case of $F$ being a function of $Z$ only, the equations
(\ref{SFBV30a}), (\ref{SFBV30b}) do provide for the exact solution
\cite{LT,BV3,BB1,BLT-EPJC,BL1},
\beq
\label{SFBV30c}
\Psi( Z, t )  =  \exp\left\{  t  \big(  E(  -  ( i \hbar
)^{ -2 } \;\! t \;\! {\rm ad}'( F ) )
( i \hbar )^{ -2 }  ( \Delta' F )  \big) ( Z )  \right\}
\exp\left\{  -  ( i \hbar )^{ -2 } \;\! t \;\! {\rm ad}'( F )  \right\}
\Psi_{ 0 }( Z ), 
\eeq
where we have denoted
\beq
\label{SFBV30d}
F  =:  F( Z ), \quad   E( X )  =:   \frac{  \exp\{ X \}  - 1 }{ X }, 
\eeq
and  $\Psi_{ 0 }( Z )$ is an initial wave function.
Provided the
first equation (\ref{SFBV30d}) holds, the $ZP$ symbol for the whole operator
(\ref{SFBV29})
corresponds to the Weyl symbol for  the second term alone in the latter operator
\cite{BL2}.

It is a remarkable feature that  the generators of the form (\ref{SFBV28})
satisfy the following composition law,
\beq
\label{SFBV31}
( i \hbar )^{ -1}  [  \mathcal{ H }_{ F } ,  \mathcal{ H }_{ F' }  ] =
\mathcal{ H }_{ F  \circ  F' },   
\eeq
where
\beq
\label{SFBV32}
F  \circ  F'   =:   ( i \hbar )^{ -2 } (  F,  F' )_{ \Delta },     
\eeq with  $( A, B )_{ \Delta }$ being the so-called quantum
$\Delta$ - antibracket \cite{BM1,BM2,K-S,Ber},
\beq
\label{SFBV33}
(A, B )_{ \Delta }  =:  \frac{1}{2}  (   [  A,  [  \Delta, B  ]  ]  -
( A \;\leftrightarrow \;  B  )
(-1)^{  ( \varepsilon_{ A }  + 1 ) ( \varepsilon_{ B } + 1 )  }  ).    
\eeq
Their main property,
\beq
\label{SFBV34}
[  \Delta,  ( A, B )_{ \Delta }  ]  =
[  [  \Delta,  A  ],  [  \Delta,  B  ]  ] ,  
\eeq
yields the (\ref{SFBV31}) immediately.
The quantum 2 - antibracket (\ref{SFBV33}) does satisfy  the modified
Jacobi relations,
\beq
\nonumber
&&(  A,  ( B,  C )_{ \Delta }  )_{ \Delta }
( -1)^{ ( \varepsilon_{ A } + 1 ) ( \varepsilon_{ C } + 1 ) }  +
  {\rm cyclic\; perm.} (  A,  B,  C )  =\\
\label{SFBV35}
&&\qquad\qquad =
  \frac{ 1 }{ 2 }  [  ( A, B, C )_{ \Delta }
(-1)^{  ( \varepsilon_{ A } + 1 )( \varepsilon_{ C } + 1 )  } ,  \Delta  ] ,    
\eeq
where the $( A, B, C )_{ \Delta }$  is the so-called quantum
3 - antibracket, and so on
\cite{BM1,BM2}.

\section{$Sp(2)$ symmetric construction}

In its $Sp(2)$ symmetric version \cite{BLT1,BLT2,BLT3,BBL},
a superfield Schroedinger equation becomes
$Sp(2)$ vector valued
\beq
\label{SFBV41}
(  i \hbar  D^{ a }  -  Q^{ a }  )  \Psi  =  0,   
\eeq where the following conventions
\footnote{For the sake of uniformity, henceforth
we make use of the notation $\Phi^{ \alpha a }$ for the former field variable
$\pi^{\alpha a}$ \cite{BLT1}. Also, as for the Boson metric $g^{ ab }$,
we assume it symmetric, constant, and
invertible, so that $g_{ ab }$ is its inverse.} hold for the required $Sp(2)$
vector valued operators
\beq
\label{SFBV42}
&& \qquad\qquad\qquad
D^{ a }  =:  \frac{ \partial }{ \partial  \tau_{ a } }  +
g^{ ab }  \tau_{ b }  \frac{ \partial }{ \partial  t } , \quad   
[  D^{ a },  D^{ b }  ]  = 2 g^{ ab } \frac{ \partial }{ \partial  t }, \\   
\label{SFBV44}
&&
\;\;\;Q^{ a }  =:   \Delta^{ a }_{ + }  -  F^{ a },  \quad   
\Delta^{ a }_{\pm}  =:  \Delta^{ a } \pm  \frac{ i }{ \hbar }  V^{ a }, \quad
F^{ a }  =:
g^{ ab } \varepsilon_{ bc }  ( i \hbar)^{ -1}  [ \Delta^{ c }_{ + } ,  B ],
\\ 
\label{SFBV46}
&&\qquad\qquad\qquad\qquad
[  \Delta^{ a },  \Delta^{ b }  ]  =  0, \quad   
\label{SFBV47}
[  \Delta^{ a }_{ \pm},  \Delta^{ b }_{\pm }  ]  =  0,\\ 
\label{SFBV48}
&&\qquad\quad Z^{ A }  =:
(  \Phi^{ \alpha }, \Phi^{ \alpha a }; \Phi^*_{ \alpha a }, \Phi^{**}_{ \alpha }),\quad
P_{ A }  =:
(  P_{ \alpha }, P_{\alpha a }; P_{ * }^{ \alpha a }, P_{ **}^{ \alpha }  ),\\
\label{SFBV48a}
&&\qquad\qquad\quad \Delta^{ a }  =:
\frac{ 1 }{ 2 }  P_{ A } E^{ AB a } P_{ B }  (-1)^{ \varepsilon_{ B } },\quad
E^{ AB a }  =  {\rm const},\\
\label{SFBV48b}
&&\qquad\qquad\qquad\qquad\quad V^{ a }  =:
-  i \hbar \;\! \varepsilon^{ ab }  \Phi^*_{\alpha b }
P_{ **}^{ \alpha }  (-1)^{ \varepsilon_{\alpha}},
\eeq
a Boson operator $B$ is restricted as to satisfy the specific  "master equation",
\beq
\label{SFBV49}
[  \Delta^{ a }_{ + },  ( B, B )^{ b }_{ \Delta_{ + } }  ]  +
(  a \leftrightarrow b  )  =  0, \quad
\quad
\varepsilon( B )  =  0,    
\eeq
with
\beq
\label{SFBV50}
(  A, B  )^{ a }_{  \Delta_{\pm}  }  =:
\frac{ 1 }{ 2 }  \left(  [  A,  [  \Delta^{ a }_{\pm },  B  ]  ]  -
(  A \;\leftrightarrow\; B)
(-1)^{  ( \varepsilon_{ A } + 1 ) ( \varepsilon_{ B } + 1 )  }\right),  
\eeq
being the $Sp(2)$ vector-valued quantum antibracket \cite{BM2}.
The main property of the quantum 2-antibracket holds,
(\ref{SFBV50}),
\beq
\label{SFBV50aa}
[  \Delta^{ a }_{\pm },  ( A, B )^{ b }_{ \Delta_{ \pm } }  ]  +  (
a \leftrightarrow b )  =
[  [ \Delta^{ a }_{ \pm },  A ],  [ \Delta^{ b }_{ \pm },  B ]  ]
+  ( a \leftrightarrow b ).  
\eeq
Also, the quantum 2 - antibracket
(\ref{SFBV50}) does satisfy the modified Jacobi relation, \beq
\nonumber &&\left(  (  A,  (  B,  C  )^{ a }_{ \Delta_{ \pm } }  )^{
b }_{ \Delta_{ \pm } } (-1)^{  ( \varepsilon_{ A }  +  1 ) (
\varepsilon_{ C }  +  1 )  }  +
{\rm cyclic\; perm}. ( A,  B,  C )  \right)  +  (  a \leftrightarrow b  )  =\\
\label{SFBV50a} &&\qquad =\frac{ 1 }{ 2 } \left(  [  ( A,  B, C )^{
a }_{ \Delta_{ \pm } } (-1)^{  ( \varepsilon_{ A }  + 1 ) (
\varepsilon_{ C }  +  1 )  }, \Delta^{ b }_{ \pm}  ]
+(  a  \leftrightarrow  b )  \right), 
\eeq
where the $( A, B, C )^{ a }_{ \Delta_{ \pm } }$ is the
so-called quantum 3 - antibracket, and so on.
In the $Sp(2)$ case, the formulae (\ref{SFBV50}), (\ref{SFBV50aa}) and (\ref{SFBV50a})
are natural counterparts
to the formulae (\ref{SFBV33}), (\ref{SFBV34}) and (\ref{SFBV35}),
respectively, in the $Sp(1)$ case.

Due to the $Sp(2)$ symmetric version of the Poincare lemma,
we have from (\ref{SFBV49})
\beq
\label{SFBV51a}
\frac{ 1 }{ 2 }  ( B, B )^{ a }_{ \Delta_{ + } }  =
( i \hbar )^{ 2 }  X^{ a }  +   i  \hbar  [  \Delta^{ a }_{ + },  Y ],   
\eeq
where $X^{ a }$ is an $Sp(2)$ vector-valued singlet Fermion operator,
\beq
\label{SFBV51b}
[ \Delta^{ a }_{ + },  X^{ b } ]  +  ( a \leftrightarrow b )  =  0,\quad
X^{ a }  \neq  [ \Delta^{ a }_{ + },  {\rm anything}  ],  
\eeq
$Y$  is an arbitrary $Sp(2)$ invariant Boson operator, "anything"
is an arbitrary $Sp(2)$ invariant Boson operator.
In the $Sp(2)$ case, the equations (\ref{SFBV49}), (\ref{SFBV51a})
are natural counterparts to the respective equations (\ref{SFBVc}),
(\ref{SFBV7d}) in the $Sp(1)$ case.

Due to the property (\ref{SFBV42}), it follows from the generating
equations  (\ref{SFBV41}),
\beq
\label{SFBV51}
i \hbar \frac{\partial } {\partial t }  \Psi  =  \mathcal{ H }  \Psi,     
\eeq
where the Hamiltonian has the well-known form commuting certainly with
the operators $\Delta^{ a }_{ + }$,
\beq
\label{SFBV52}
\mathcal{ H }  =:
-  \frac{ 1 }{ 4 }  g_{ ab}  ( i \hbar )^{ -1}  [  Q^{ a },  Q^{ b }  ]  =
\frac{ 1 }{ 2 }  ( i \hbar )^{ -2}
[ \Delta^{ a }_{ + }, \varepsilon_{ ab } [ \Delta^{ b }_{ + },  B  ]  ].  
\eeq
The terms quadratic in $B$ in the $\mathcal{ H }$ drop out as follows,
\beq
\nonumber
&&-  \frac{ 1 }{ 4 }  g_{ ab }  ( i \hbar )^{-1}  [ F^{ a },  F^{ b } ]  =
-  \frac{ 1 }{ 4 }  g^{ ab }  \varepsilon _{ ac }  \varepsilon_{ bd }
( i \hbar )^{ -3 } [  [  \Delta^{ c }_{ + },  B  ],
[  \Delta^{ d }_{ + },  B  ]  ]  =\\
\label{SFBV53}
&&\qquad\qquad = -  \frac{ 1 }{ 8 }  g^{ ab }   \varepsilon_{ ac }  \varepsilon_{ bd }
( i \hbar )^{ -3 } \left(  [ \Delta^{ c }_{ + },
( B, B )^{ d }_{ \Delta_{ + } }]  +  ( c \leftrightarrow d )  \right)   =  0.   
\eeq
Here in the (\ref{SFBV53}),
in  the last equality, we have used the (\ref{SFBV50aa}) and then the (\ref{SFBV49}).
The superfield $\Psi$  has the component form
\beq
\label{SFBV54}
\Psi( t, \tau, Z )  =
\exp\left\{  \tau_{ a }  ( i \hbar )^{-1}  Q^{ a }  \right\}
\Psi_{ 0 }( t, Z ),  
\eeq
where the zero component satisfies by itself
the Schroedinger equation (\ref{SFBV51})
with the Hamiltonian (\ref{SFBV52}).
The same as in the $Sp(1)$ case,
the arbitrariness in a solution to the  quantum master
equations
\beq
\label{SFBV55}
\Delta^{ a }_{ + }  \Psi_{ 0 }  =  0,  \quad \varepsilon(\Psi_{ 0})=0,\quad
\Psi_{ 0 }  =  \exp\left\{  \frac{ i }{ \hbar }  W  \right\},   
\eeq
is measured by the evolution operator with the Hamiltonian (\ref{SFBV52}).

It seems a bit strange that the Boson $B$ is restricted as to satisfy
the equations (\ref{SFBV49}), although
the standard expression in the right-hand side of the second equality in
(\ref{SFBV52})
does commute
with the $\Delta^{ a }_{ + }$ as for an arbitrary $B$.
The reason is just the second equality (\ref{SFBV52}) by
itself.  In order to clarify the matter, let us consider
the definition of the Hamiltonian $\mathcal{ H }$
in  a natural basis,
\beq
\label{SFBV56}
g^{ ab }  =  g_{ ab }  =: \begin{pmatrix}0&1\\1&0\end{pmatrix}, 
\quad
\varepsilon^{ ab }  =  -  \varepsilon_{ ab }  =:
\begin{pmatrix}0&1\\-1&0\end{pmatrix},   
\eeq
so that
\beq
\label{SFBV57}
D^{ 1 }  =  \frac{ \partial }{ \partial  \tau_{ 1 } }  +
  \tau_{ 2 }  \frac{ \partial }{ \partial  t },\quad
D^{ 2 }  =  \frac{ \partial }{ \partial  \tau_{ 2 } }  +
  \tau_{ 1 }  \frac{ \partial }{ \partial  t },\quad
  g^{ ab } \varepsilon_{ bc }  = \begin{pmatrix}1&0\\0&-1\end{pmatrix}.
\eeq
First of all, we have, for the Hamiltonian $\mathcal{ H }$, the
first equation in (\ref{SFBV52}),
\beq \label{SFBV58} \mathcal{ H }  =  -
\frac{ 1 }{ 2 }  ( i \hbar )^{-1}  [ Q^{ 1 }, Q^{ 2 } ],     
\eeq
where
\beq
\label{SFBV59}
Q^{ 1 }  =  \Delta^{ 1 }_{ + }  -  F^{ 1 },  \quad
  Q^{ 2 }  =  \Delta^{ 2 }_{ + }  -  F^{ 2 },   
\eeq
\beq
\label{SFBV60}
F^{ 1 }  =  ( i \hbar )^{-1}  [  \Delta^{ 1 }_{ + },  B  ],\quad
F^{ 2 }  = -  ( i \hbar )^{-1}  [  \Delta^{ 2 }_{ + },  B  ],   
\eeq
so that
\beq
\nonumber
\mathcal{ H }  &=&  -  \frac{ 1 } {2 }  ( i \hbar )^{-2}
\big(  [  \Delta^{ 1 }_{ + },  [  \Delta^{ 2 }_{ + },  B  ]  ]   -
(  1 \leftrightarrow 2 )    -\\
\label{SFBV61}
&&\quad\!-
 ( i \hbar )^{-1}
 [  [ \Delta^{ 1 }_{ + },  B  ],  [ \Delta^{ 2 }_{ + },  B  ]  ]  \big). 
\eeq
In order to provide for the operators $Q^{ 1 }$ and $Q^{ 2 }$, ( \ref{SFBV59}),
to commute with the  Hamiltonian
$\mathcal{ H }$, (\ref{SFBV58}), both the charges (\ref{SFBV59}) should be nilpotent,
\beq
\label{SFBV62}
[  [ \Delta^{ 1 }_{ + }, B ],  [ \Delta^{ 1 }_{ + }, B ]  ]  =  0,  \quad
  [  [ \Delta^{ 2 }_{ + }, B ],  [ \Delta^{ 2 }_{ + }, B ]  ]  =  0.   
\eeq
The first and the second equations in (\ref{SFBV62})
are exactly the equations (\ref{SFBV49})
at $a = b = 1$ and at $a = b = 2$,
respectively. Now, in the first line in the right-hand side in (\ref{SFBV61})
we recognize exactly the standard
expression in the right-hand side in the second equality in (\ref{SFBV52}).
In turn, the equation (\ref{SFBV49}) at $a = 1, b = 2$,
or vice versa, cancels the expression in the second line in (\ref{SFBV61}).
Thus, we have explained in
detail how  the equations (\ref{SFBV49}) for the Boson operator $B$
do come from the general structure (\ref{SFBV58})
of the  Hamiltonian $\mathcal{ H }$ as constructed of
the two nilpotent supercharges $Q^{ 1 }$ and $Q^{ 2 }$.
In contrast to the $Sp(1)$ case, in the $Sp(2)$ symmetric
superfield formalism, the  equations (\ref{SFBV49})
are just the price of the higher supersymmetry.

Finally, consider, in the $Sp(2)$ case, the composition law similar
to the one of (\ref{SFBV31}) and (\ref{SFBV32}),
as for the Hamiltonian (\ref{SFBV52}) rewritten as
\beq
\label{SFBV63}
\mathcal{ H }_{ F^{ 2} }  =
( i \hbar )^{-1}  [  \Delta^{ 1 }_{ + },  F^{ 2 }  ],   
\eeq
where $F^{ 2 }$ is given by the second in (\ref{SFBV60}).
Then, the composition law has just the
form (\ref{SFBV31}), (\ref{SFBV32}), with the $\Delta^{ 1 }_{ + }$
and the $F^{ 2 }$
standing for the $\Delta$ and
the $F$, respectively. Vice versa, we could make use of the $\Delta^{ 2 }_{+}$ and
the $F^{ 1 }$
as to stand for the $\Delta$ and the $F$, respectively, when having
the Hamiltonian (\ref{SFBV52})
rewritten equivalently as
\beq
\label{SFBV64}
\mathcal{ H }_{ F^{1} }  =
( i \hbar )^{-1}  [  \Delta^{ 2 }_{ + },  F^{ 1 }  ],     
\eeq
where $F^{ 1 }$ is given by the first in (\ref{SFBV60}).

\section{General nilpotency}

Here, we present in both the $Sp(1)$ and the $Sp(2)$ cases, in parallel,
the simplest class
of  solutions for the Hamiltonian.
In the $Sp(1)$ case, we strengthen the (\ref{SFBV7d})
to the nilpotency condition for the Fermion $F$,
\beq
\label{SFBV66}
\mathcal{ H }_{ S }  =  0,  \quad  G  =  0,  \quad
\Rightarrow \quad  [  F,  F  ]  =  0.    
\eeq
Then, we have for the Hamiltonian,
\beq
\label{SFBV67}
\mathcal{ H }  =  ( i \hbar )^{-1}  [ \Delta,  F ].  
\eeq
In the case of being the $F$ a function of  $Z^{ A }$ only,
the condition (\ref{SFBV66}) is satisfied
automatically.
In the $Sp(2)$ case, we strengthen the (\ref{SFBV51a})
to the "nilpotency" condition for the Boson $B$,
\beq
\label{SFBV68}
X^{ a }  =  0, \quad   Y  =  0, \quad\Rightarrow    \quad
(  B,  B  )^{ a }_{ \Delta_{ + } }  =  0.  
\eeq
Then, we have for the Hamiltonian,
\beq
\label{SFBV69}
\mathcal{ H }  =  \frac{ 1 }{ 2 }  ( i \hbar )^{ -2 }
[  \Delta^{ a }_{ + },  \varepsilon_{ ab }  [  \Delta^{ b }_{ + },  B  ]  ].  
\eeq
In the case of being the $B$ a function of fields only, the equation (\ref{SFBV68})
is satisfied automatically.

\section{ Heisenberg equations of motion in terms of quantum
antibrackets}

Here, we present in both the $Sp( 1 )$ and the $Sp( 2 )$ cases, in parallel, the Heisenberg
equations of motion in terms of the quantum antibrackets.
Denote by $\Gamma$ the full set of the Schroedinger canonical variable  operators,
\beq
\label{SFBV70}
\Gamma  =:  (  Z^{ A } ;  P_{ A }  ),    
\eeq
and let $\tilde{ \Gamma }( t , \tau )$  be the respective superfield Heisenberg
canonical variable operators.

In the $Sp( 1 )$ case, the superfield Heisenberg equations of motion have the form,
\beq
\label{SFBV71}
i \hbar  D  \tilde{ \Gamma }  =  [\tilde{ Q },  \tilde{\Gamma}] , \quad 
i \hbar  D  \tilde{ Q }  =  [  \tilde{ Q },  \tilde{ Q }  ].    
\eeq
It follows from these equations that \cite{BM2},
\beq
\label{SFBV73}
( i \hbar )^{ 2 }  \frac{ \partial }{ \partial t }  \tilde{ \Gamma }  =
 -  \frac{ 1 }{ 2 }  [  \tilde{ \Gamma },  [  \tilde{ Q },  \tilde{ Q }  ]  ]  =
 -  \frac{ 2 }{ 3 }  (  \tilde{ \Gamma },  \tilde{ Q }  )_{ \tilde{ Q } },    
\eeq
where the quantum 2 - antibracket, $(A,  B)_{ Q }$, is
defined by the (\ref{SFBV33}), with $Q$, the third in the
(\ref{SFBV2}),  standing for the $\Delta$.

In the $Sp( 2 )$ case,  the respective superfield Heisenberg
equations of motion have the form,
\beq
\label{SFBV74}
i \hbar  D^{ a }  \tilde{ \Gamma }  =  [  \tilde{ Q }^{ a },  \tilde{ \Gamma }  ], \quad 
i \hbar  D^{ a }  \tilde{ Q }^{ b }  =  [  \tilde{ Q }^{ a },  \tilde{ Q }^{ b }  ].   
\eeq
It follows from these equations that
\beq
\label{SFBV76}
( i \hbar )^{ 2 }  \frac{ \partial }{ \partial t }  \tilde{ \Gamma }  =
 -  \frac{ 1 }{ 4 }  g_{ ab }  [  \tilde{ \Gamma },
[  \tilde{ Q }^{ b }, \tilde{ Q }^{ a }  ]  ]  =
 -  \frac{ 1 }{ 3 }  g_{ ab }
(  \tilde{ \Gamma },  \tilde{ Q }^{ b }  )^{ a }_{ \tilde{ Q } },  
\eeq
where the $Sp( 2 )$ vector valued  quantum 2 - antibracket,  $(  A,  B  )^{ a }_{ Q }$, is defined
by the (\ref{SFBV50}), with $Q^{ a }$, the first in the (\ref{SFBV44}),
standing for the $\Delta^{ a }_{\pm}$.

\section{Conclusion}
\noindent
In the present paper, within the superfield approach, we have proposed the new quantum
generating equation (\ref{SFBV2}) for the general field-antifield formalism.
The three basic Fermion
objects,  the super-time covariant derivative $D$,  the odd Laplacian $\Delta$,
and the hyper-gauge  Fermion $F$,  enter that linear homogeneous generating equation,
in a quite symmetric
way.  Then,  from the generating equation,
we have derived the Schroedinger equation (\ref{SFBV5})
with  the Hamiltonian $\mathcal{ H }$, (\ref{SFBV6}), commuting with the
supercharge Q, the third in
(\ref{SFBV2}). It follows from the latter property (\ref{SFBV7a})
that the Hamiltonian $\mathcal H$
commutes with the $\Delta$,
provided the $\mathcal{ H }$ commutes with the $F$, as well.
Thus, we have  determined
the general structure (\ref{SFBV7f})
of the Hamiltonian (\ref{SFBV6}). As usual, the Hamiltonian consists
of a singlet component and a
$\Delta$-exact component.  We have shown that the $\Delta$-exact components
(\ref{SFBV28})
serve as
generators  to the quantum master-transformations.  In turn,
we have shown that these
generators (\ref{SFBV29})  do satisfy the nice composition law (\ref{SFBV31})
given by (\ref{SFBV32}) in terms of the
quantum antibrackets  (\ref{SFBV33}).
We have also presented an $Sp(2)$ symmetric extension to the main construction,
with specific features caused by the principal fact that all basic equations become
$Sp(2)$ vector-valued ones.
\\

\section*{Acknowledgments}
\noindent The authors  would like  to thank Klaus Bering of Masaryk
University for interesting discussions. The work of I. A. Batalin is
supported in part by the RFBR grants 17-01-00429 and 17-02-00317.
The work of P. M. Lavrov is supported in part by  the RFBR grant
16-52-12012-NNIO.

\end{document}